\documentclass[12pt]{article}

\usepackage[T1]{fontenc}
\usepackage{amsmath, amsthm, amsfonts, amssymb}
\usepackage[margin=1in]{geometry}
\usepackage{graphicx, wrapfig, subfig}
\usepackage{setspace}
\usepackage{authblk}
\usepackage{breakcites} 
\usepackage{fixltx2e}
\usepackage{multirow}
\usepackage{booktabs}
\usepackage{hyperref}

\graphicspath{ {./figs/} }
\def\bs{\boldsymbol}

\usepackage[margin=1in]{geometry}
\usepackage{lineno}
\usepackage[numbers,super]{natbib}
\usepackage{url}
\makeatletter
\renewcommand\@biblabel[1]{#1.}
\makeatother

\begin{document}

\title{Data-augmented modeling of intracranial pressure}

\author[1,2]{Jian-Xun Wang\protect\thanks{Email address: \href{mailto:jwang33@nd.edu}{\protect\nolinkurl{jwang33@nd.edu}}}}
\author[3]{Xiao Hu\protect\thanks{Email address: \href{mailto:xiao.hu@ucsf.edu}{\protect\nolinkurl{xiao.hu@ucsf.edu}}}}

\author[1]{Shawn C. Shadden\protect\thanks{Email address: \href{mailto:shadden@berkeley.edu}{\protect\nolinkurl{shadden@berkeley.edu}}}}

\affil[1]{\small\textit{Mechanical Engineering}, \small\textit{University of California}, \small\textit{Berkeley, CA}}
\affil[2]{\small\textit{Aerospace and Mechanical Engineering, Center of Informatics and Computational Science}, \small\textit{University of Notre Dame}, \small\textit{Notre Dame, IN}}
\affil[3]{\small\textit{Department of Physiological Nursing, Department of Neurological surgery, Institute of Computational Health Sciences, UCSF Joint Bio-Engineering Graduate Program}, \small\textit{University of California}, \small\textit{San Francisco, CA}}

\date{}

\maketitle

\begin{description}
    \item[abbreviated title:]
        Data-augmented modeling of ICP
    \item[correspondence:]
        Jian-Xun Wang,
        Aerospace and Mechanical Engineering, Center of Informatics and Computational Science, University of Notre Dame, Notre Dame, IN, USA.
        e-mail: \texttt{jwang33@nd.edu}
\end{description}

\begin{abstract}
Precise management of patients with cerebral diseases often requires intracranial pressure (ICP) monitoring, which is highly invasive and requires a specialized ICU setting. The ability to noninvasively estimate ICP is highly compelling as an alternative to, or screening for, invasive ICP measurement. Most existing approaches for noninvasive ICP estimation aim to build a regression function that maps noninvasive measurements to an ICP estimate using statistical learning techniques. These data-based approaches have met limited success, likely because the amount of training data needed is onerous for this complex applications. In this work, we discuss an alternative strategy that aims to better utilize noninvasive measurement data by leveraging mechanistic understanding of physiology. Specifically, we developed a Bayesian framework that combines a multiscale model of intracranial physiology with noninvasive measurements of cerebral blood flow using transcranial Doppler. Virtual experiments with synthetic data are conducted to verify and analyze the proposed framework. A preliminary clinical application study on two patients is also performed in which we demonstrate the ability of this method to improve ICP prediction. 
\end{abstract}

\textbf{keywords:}  cerebrovascular dynamics; data assimilation; patient-specific modeling; transcranial Doppler.

\newpage

\section{Introduction}
\label{sec:intro}
Determination of intracranial pressure (ICP) is essential for precise management of patients with brain injury, hemorrhage, tumor, hydrocephalus and other neurologic conditions~\cite{andrews2008nicem}. Elevated ICP reduces cerebral blood flow, which can lead to brain damage or death~\cite{ghajar2000traumatic}. The clinical standard for ICP monitoring, which entails penetrations of the skull and brain, carries the risks of hemorrhage, infection and tissue damage~\cite{hickman1990intracranial}. Moreover, such invasive techniques require neurosurgical expertise and a specialized ICU setting~\cite{zhang2017invasive}. Even in such a setting, a significant concern is to identify when ICP monitoring should be initiated for a given patient. A noninvasive method to estimate ICP can reduce the risks of invasive ICP monitoring, better identify patients needing invasive monitoring, and potentially broaden ICP evaluation beyond the ICU setting. 

The majority of noninvasive ICP (nICP) research has been to identify noninvasive signals that are correlated to ICP. These have included pupil size, intraocular pressure, optic nerve sheath diameter, tympanic membrane displacement, cerebral blood flow velocity (CBFV), visual evoked potentials and skull movements, among others~\cite{asiedu2014review}. However, identifying noninvasive signals that are correlated to ICP often only enables inference of ICP trending or its detrimental effects. There remains a need to quantitatively estimate ICP from noninvasive signals for proper clinical response~\cite{zhang2017invasive, asiedu2014review}. To address this challenge, nICP research has recently sought to develop algorithmic solutions that can bridge the gap between noninvasive measurements (measurable states) and ICP (the hidden state)~\cite{cardim2016non}. 

To connect measurable states with a hidden state requires a model, which can be data-based or theory-based. Most prior works on nICP estimation have been data-based, and have tried to construct mapping functions between noninvasive signals and ICP using supervised learning techniques, including linear/nonlinear regression~\cite{brandi2010transcranial}, support vector machines (SVM)~\cite{xu2010improved}, kernel spectral regression~\cite{kim2012noninvasive}, and artificial neural networks~\cite{hu2006data}. Despite the varied attempts, these methods have struggled to achieve accurate nICP assessment for a \emph{de novo} patient. The limitation of a data-based approach is the requirement of sufficient training data. Data is inherently limited for this problem because gold-standard ICP measurement is highly invasive, data can vary in quality or consistency, and complications with sharing patient data. Moreover, a large amount of data is likely necessary due to the complexity of the underlying physiology and inter-patient variability. 

Utilization of {\em theory-based models} may help to alleviate the need for inordinate training data, and maximize the utility of each individual's data, when compared to a data-based approach. Theory-based intracranial modeling has advanced in recent years to increase our understanding of the mechanisms that drive intracranial pressure~\cite{wakeland2008review,linninger2016cerebrospinal}. In contrast to black-box models that depend on training data, theory-based models rely on physiological knowledge and physical principles. It is broadly accepted that ICP dynamics is driven by the interactions between cerebral blood flow (CBF), cerebrospinal fluid (CSF), and brain soft tissue under the constraints of a rigid skull. Lumped-parameter (LP) models are widely employed for modeling the dynamics of these intracranial components, and among the several publications in this area, Ursino et al.~\cite{ursino1998interaction,ursino2010model}, Stevens et al.~\cite{stevens2005modeling}, and Linninger et al.~\cite{linninger2009mathematical} have contributed significantly to establishing theoretical models of the component dynamics. However, the clinical impact of existing theory-based models remains negligible. 

The challenges of using theory-based model for ICP estimation include the coupling of sufficiently comprehensive component models needed to capture the important physiology, and calibration of these model parameters for a \emph{de novo} patient. A promising approach is to combine useful information from both theory-based modeling and noninvasive measurement. Kashif et al.~\cite{kashif2012model} demonstrated the merits of this idea. Namely, they showed that the accuracy of a model-based nICP approach was significantly improved compared to a purely data-driven approach. Hu et al.~\cite{hu2007estimation} also exploited using a basic physiologic model with measured data and filtering to estimate ICP. A recent review~\cite{cardim2016non} comprehensively comparing existing nICP algorithms also confirmed the advantage of introducing physical models/constraints. While these works substantiate the potential of this approach, theory-based nICP methods still require significant development both in terms of modeling the physiology and effective assimilation of data. Data assimilation (DA) is emerging in other areas of biomechanics modeling~\cite{marsden2014optimization}, and has recently included the use of variational-based methods \cite{tiago2017velocity, itu2017personalized}, unscented Kalman filtering~\cite{bertoglio2012sequential, moireau2013sequential, pant2016data}, and most recently ensemble Kalman filtering (EnKF)~\cite{arnold2017uncertainty, lal2017non}.

The framework developed herein advances both intracranial modeling capabilities and data assimilation methodology in comparison to prior works in nICP estimation. Namely, we employ a Bayesian data assimilation (DA) framework that uses a regularizing iterative ensemble Kalman filtering to combine noninvasive transcranial Doppler (TCD) measurements with a recent multiscale intracranial dynamics model. The novelty of this work is in the state-of-the-art Bayesian DA and intracranial dynamics modeling, as well as an alternative from existing data-based, black-box nICP methods. Moreover, this work is significant in that the performance of the proposed approach in both synthetic and patient-specific cases demonstrates that TCD CBF measurements are informative of ICP dynamics, and that ICP can be potentially estimated noninvasively from CBF waveforms. The rest of this paper is organized as follows. Section~\ref{sec:meth} introduces the key components of the proposed model-based nICP framework, including the multiscale intracranial model and regularizing iterative ensemble Kalman method. Section~\ref{sec:result} presents numerical results for both synthetic cases and patient-specific cases to demonstrate merits of the proposed method. Finally, the success and limitations of the method are discussed in Section~\ref{sec:dis}.
 
\section{Materials and Methods}
\label{sec:meth}
\subsection{Overview of data-augmented, theory-based modeling framework}
\label{subsec:overview}
The main idea of the proposed framework is to combine a physiological model of ICP dynamics and noninvasive ICP-related measurements (e.g., CBFV or arterial blood pressure, ABP) to achieve an nICP estimation. A Bayesian data assimilation scheme is adopted to incorporate the noninvasive data for calibrating the model and estimating unobserved states (i.e., ICP) for a \emph{de novo} patient.
\begin{figure}[!htbp]
	\centering
	\includegraphics[width=0.8\textwidth]{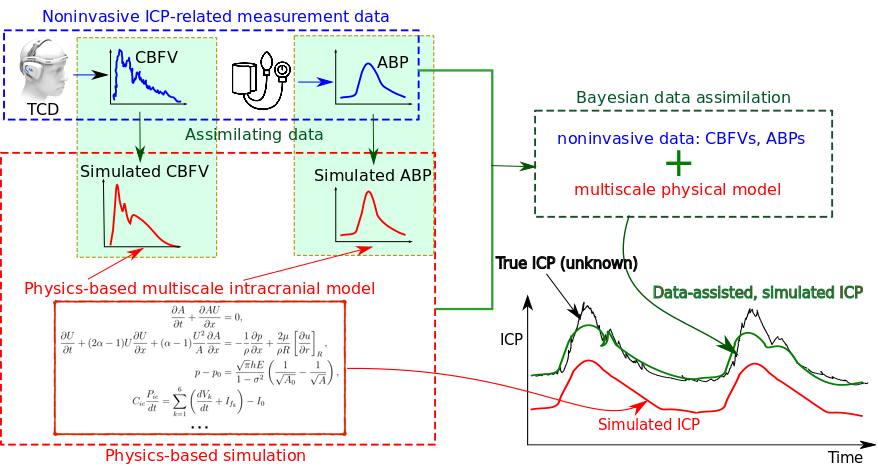}
	\caption{Schematic of the proposed data-augmented, theory-based framework for ICP dynamics. By assimilating noninvasively measurable data (e.g., CBFV and/or ABP at certain vessels) into the theory-based physiological model, predictions of the unobservable states (e.g., ICP) can be significantly improved.}
	\label{fig:idea}
\end{figure}
A schematic of this framework is shown in Fig.~\ref{fig:idea}. Conceptually, it consists of three modules, including (1) a forward model of intracranial dynamics, (2) noninvasive measurement data, and (3) a data assimilation scheme, which are marked by the red, blue and green boxes, respectively, in Fig.~\ref{fig:idea}. The theory-based forward model within the red box is used to compute measurable states (e.g., CBFV and ABP) and hidden states (e.g., ICP) based on physical principles. Without calibration, the model can be expected to produce an inaccurate prediction (red curve) due in large part to inaccurate model parameters for a {\em de novo} patient. To address this issue, noninvasive measurement data specific to each patient are integrated into the model within a Bayesian framework, and thus primary model parameters can be more accurately established, leading to improvement in the calibrated ICP prediction (green curve). 

Specifically, a multiscale cerebrovascular model~\cite{ryu2015coupled} is employed as the forward model in this work to simulate intracranial states $\mathbf{z}$ (e.g., ICP, CBFV, and ABP) based on prescribed initial states $\mathbf{z}_{0}$, boundary conditions $\partial D$, and model parameters $\bs{\theta}$. The forward problem can be generally represented as a mapping $\mathbf{z} = \mathcal{F}(\mathbf{z}_{0}, \partial D, \bs{\theta})$. The predicted state $\mathbf{z}$ can be expected to be biased from the truth $\mathbf{\tilde{z}}$ due to model-form errors in $\mathcal{F}$ and uncertainties in boundary conditions $\partial D$ and parameters $\bs{\theta}$. This can be expressed as 
\begin{equation}
\label{eq:prior}
\mathbf{z} = \mathcal{F}(\mathbf{z}_{0}, \partial D, \bs{\theta}) = \tilde{\mathbf{z}} + \bs{\sigma}_m,
\end{equation}
where $\bs{\sigma}_m$ represents the model discrepancy. Similarly, noninvasive measurement data $\mathbf{y}$, e.g., CBFV or systemic ABP,  to be assimilated are imprecise and indirect in relation to ICP.  This can be expressed as 
\begin{equation}
\label{eq:data}
\mathbf{y} = \mathcal{H}(\tilde{\mathbf{z}}) + \bs{\sigma}_d,
\end{equation}
where $\mathcal{H}(\cdot)$ represents a projection operator mapping the full state to the observed space, and $\bs{\sigma}_d$ represents measurement noise. Typically, the measurements are sparse in time and/or space. In the proposed framework, the model discrepancy $\bs{\sigma}_m$ is modeled as a random process representing an epistemic uncertainty, while the data noises are modeled as independent Gaussian random variables. The fusion of the model and data are formulated in a Bayesian manner. Namely, the \emph{prior} estimation is obtained from the baseline model by assuming prior distributions for the initial conditions $\mathbf{z}_o$, boundary conditions $\partial D$, and parameters $\bs{\theta}$. The \emph{likelihood} is obtained from the probabilistic distribution of the data uncertainty, and the data-assimilated prediction is the \emph{posterior} estimation obtained after the Bayesian updating. 

The forward model and data assimilation scheme are described further below, as well as the assimilation of noninvasive data from both synthetic experiments and actual patient-specific scenarios. 

\subsection{Forward model of intracranial dynamics}
\begin{figure}[h]
	\centering 
	\includegraphics[width=0.85\textwidth]{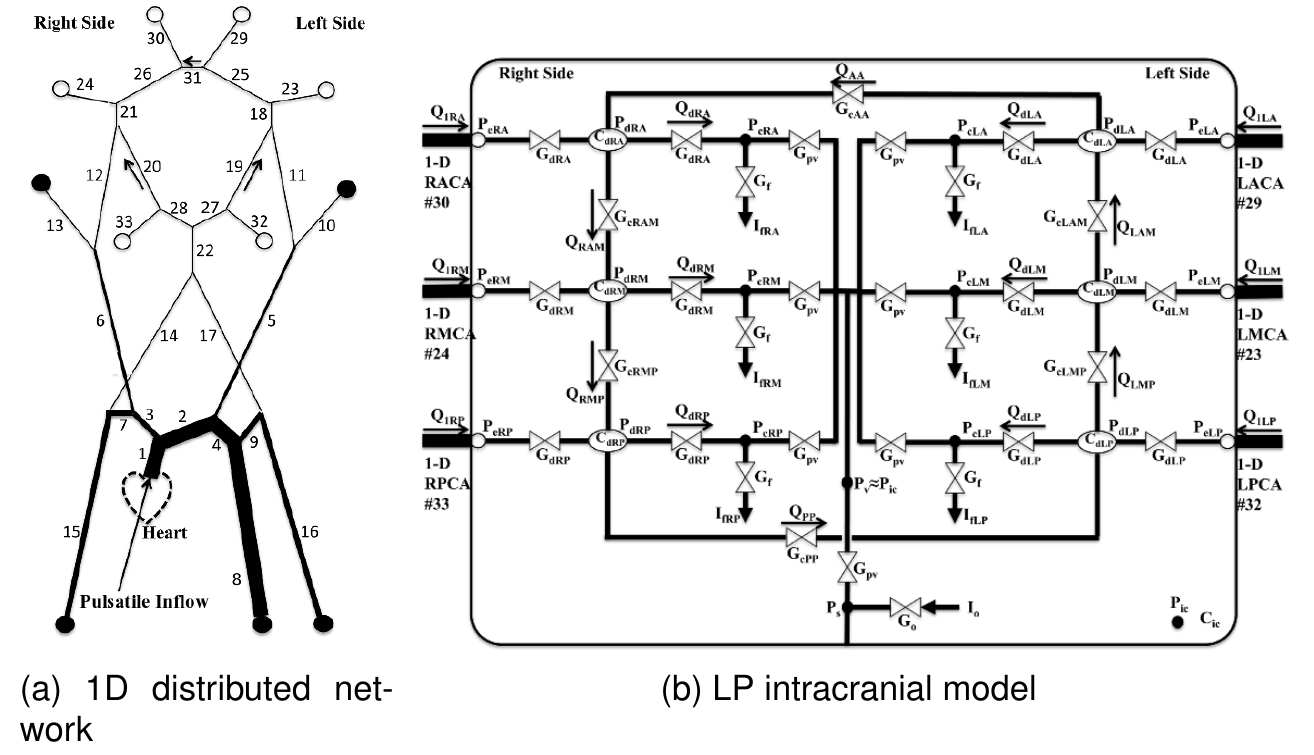}
	\caption{Schematic of the multiscale cerebrovascular model coupling (a) a distributed 1D propagation network model for major systemic arteries and (b) a lumped parameter (LP) network for intracranial dynamics. Outflow in the 1D portion marked with open circles are coupled with the LP network in (right), and boundaries marked with closed circles are coupled to 3-element Windkessel models. The bounding box represents intracranial space. Adapted from~\cite{ryu2015coupled}.}
	\label{fig:1DLP}
\end{figure}

The multiscale cerebrovascular model described in Ryu et al.~\cite{ryu2015coupled} was adopted as the forward model. This model was developed to simulate regulatory cerebrovascular flow by coupling a distributed one-dimensional (1D) propagation network model of the major systemic arteries to a sophisticated lumped parameter (LP) network of the intracranial dynamics. The intracranial LP portion of the model includes mechanisms such as cerebral autoregulation, collateral rerouting, and CSF and ICP coupling. A schematic of the multiscale forward model is shown in Fig.~\ref{fig:1DLP}. 

The 1D distributed network (Fig.~\ref{fig:1DLP}a) models arterial blood flow and pressure throughout the major systemic and cerebral arteries. While the number of the arteries is adjustable, we included the major arteries supplying the head, arms,  and torso as shown in Fig.~\ref{fig:1DLP}a. Each arterial segment is modeled as a deformable tube with blood flow and wall deformation governed by the 1D Navier-Stokes and Laplace equations,
\begin{subequations}
	\label{eq:1D-NS} 
	\begin{align}
	\frac{\partial A}{\partial t} + \frac{\partial AU}{\partial x} & = 0,\\
	\frac{\partial U}{\partial t} + (2\alpha-1)U\frac{\partial U}{\partial x} + (\alpha - 1)\frac{U^2}{A}\frac{\partial A}{\partial x} & = -\frac{1}{\rho}\frac{\partial p}{\partial x} + \frac{2\mu}{\rho R}\left[\frac{\partial u}{\partial r}\right]_R, \\
	p-p_0 & = \frac{\sqrt{\pi}h E}{1-\sigma^2} \left( \frac{1}{\sqrt{A_0}}-\frac{1}{\sqrt{A}} \right),
	\end{align}
\end{subequations} 
where $x$ and $r$ are the axial and radial coordinates, $R$, $h$, $A$, $\sigma$ and $E$ are respectively the vessel radius, thickness, cross-section area, Poisson's ratio and Young's modulus. $U$ is the transverse average of the axial velocity $u$, $p$ is the transversely averaged pressure, $\rho$ and $\mu$ are blood density and viscosity, $p_0$ is the external pressure, and $R_0$ is the radius at zero transmural pressure ($p=p_0$). The parameter $\alpha$ and wall shear rate $[\frac{\partial u}{\partial r}]_R$ are determined from an assumed velocity profile~\cite{hughes1973one} with the arterial segment diameter. In regards to boundary conditions, a sinusoidal inflow rate $Q_{in}(t)$ is prescribed at the aortic root, the extracranial terminals (marked with $\bullet$ in Fig.~\ref{fig:1DLP}a) are coupled to three-element Windkessel models, and the intracranial terminals (marked with $\circ$ in Fig.~\ref{fig:1DLP}a) are coupled with the LP intracranial network. 

An intracranial LP network (Fig.~\ref{fig:1DLP}b) is coupled to the 1D domain to capture the dynamics and coupling between CBF, CSF and ICP.  The 6 major arterial territories of the brain (Left/Right Anterior/Middle/Posterior) are represented by lumped vessel models, and are controlled by the respective vascular passive elastic tension $T_e$, viscous tension $T_{\nu}$, and active tension $T_m$ produced by the smooth muscle contraction in response to autoregulation stimuli--either myogenic or metabolic. Briefly, the relation between transmural pressure and wall tensions is applied based on Laplace's Law,
\begin{equation}
P_dr_d - P_{ic}(r_d + h_d) = T_e + T_{\nu} + T_m,
\end{equation}
where $P_d$, $r_d$, and $h_d$ are pressure, effective radius, and vessel thickness of each lumped arterial bed, and $P_{ic}$ represents the intracranial pressure. The passive elastic tension is calculated by assuming an exponential functional form of $r_d$ as,
\begin{equation}
T_e = \left[\sigma_{e0}\left(\exp(K_{\sigma}\frac{r_d - r_{d0}}{r_{d0}}) - 1\right) - \sigma_{coll}\right]h_d,
\end{equation}
where $\sigma_{e0}$, $r_{d0}$, $K_{\sigma}$, and $\sigma_{coll}$ are constant model parameters. The viscous tension is related to the viscous force of the blood flow, which is expressed as $T_{\nu} = (\eta/r_{\nu0})(dr_d/dt)$ with $\eta$ and $r_{\nu0}$ being constant model parameters. Cerebral autoregulation is carried by smooth muscle producing an elastic tension as,
\begin{equation}
T_m = T_0(1 + M)\exp\left(-\left|\frac{r_d - r_m}{r_t - r_m}\right|^{n_m}\right)
\end{equation}
where $T_0$, $r_t$, $r_m$, and $n_m$ are constant model parameters, and $M$ is the autoregulation activation factor responding to maintain CBF, which varies between [-1, 1] and can be calculated by,
\begin{equation}
M = \frac{e^{2x} - 1}{e^{2x} + 1}.
\end{equation}
The extreme values of $M$, 1 and -1, represent maximal vasoconstriction and vasodilation. To maintain CBF, the control function is modeled with a first-order low pass system expressed as,
\begin{equation}
\label{eq:qn} 
t_{CA}\frac{dx}{dt} = -x + G_{CA}\frac{q_d - q_n}{q_n},
\end{equation}
where $q_d$ is the CBF at each cerebral territory, and $q_n$ is the respect target flow rate. $t_{CA}$ and $G_{CA}$ are constant parameters representing the time scale and gain of the low pass filter, respectively.

ICP $P_{ic}$ is spatially uniform within the intracranial compartment and shared by the six distal vascular beds. The ICP and its coupling with the cerebral vascular system are determined by the Monro-Kellie principle, assuming that the total volume inside the cranium remains constant, which can be represented as follows,
\begin{equation}
\label{eq:pressureBalance} 
C_{ic}\frac{P_{ic}}{dt} = \sum_{k=1}^{6}\left(\frac{dV_k}{dt} + I_{f_k}\right) - I_0
\end{equation}
where $k$ represents the indices of six distal vascular beds, $V_k$ is the blood volume of vascular bed $k$, $I_{f_k}$ and $I_0$ are CSF inflow and outflow, respectively. The intracranial compliance $C_{ic}$ is modeled as a nonlinear function of ICP. The volume changes of $V_k$ is represented by a differential equation of effective vessel radius $r_d$ of each vascular territory, which varies due to blood pressure and myogenic or metabolic autoregulation. This cerebrovascular model has been validated against clinical measurements of a transient hyperemic response test~\cite{giller1991bedside}, which quantifies the dynamics change of CBFV in the right MCA due to transient compression of the carotid artery. Additionally, qualitative validation of the model with regards to $\mathrm{CO_2}$ inhalation and hyperventilation tests have also been  performed~\cite{ryu2017numerical}. Further implementation details and nominal parameter assignment for this model can be found in~\cite{ryu2015coupled}.

{This multi-scale model is potentially advantageous for several reasons. A distributed 1D network for modeling the major systemic and cerebral arteries facilitates data assimilation. Namely, measurements of blood flow or pressure from specific arteries can be more directly assimilated to corresponding locations in the model. Moreover, the 1D distributed network enables more realistic pressure and flow temporal waveforms~\cite{shi2011review}, and therefore, measurements of (e.g., CBFV or ABP) temporal waveform dynamics can be better assimilated, potentially better informing model calibration and ICP estimation. These pressure and flow waveforms are also the main ``forcing functions'' to intracranial dynamics. The multi-scale model also enables more avenues to make the model patient-specific from, e.g., angiography, or other clinically-available data.}

\subsection{Regularizing iterative ensemble Kalman method}

Data-assisted predictions of unobserved states and parameters can be considered posterior estimations calculated from the prior (Eq.~\ref{eq:prior}) and data (Eq.~\ref{eq:data}) using Bayes' theorem 
\begin{equation}
p(\mathbf{x}|\mathbf{y}) \sim p(\mathbf{x})p(\mathbf{y}|\mathbf{x}),
\end{equation}
where $p(\mathbf{x}|\mathbf{y})$, $p(\mathbf{x})$, and $p(\mathbf{y}|\mathbf{x})$ are the probability density functions of the posterior state, prior state, and data uncertainty, respectively. To obtain the exact posterior estimation, Markov chain Monte Carlo (MCMC) sampling is typically required to sample the posterior distribution. This process involves an onerous number of forward model evaluations sequentially, which is prohibitively expensive for nontrivial systems. As such, we adopt an approximate Bayesian approach, the iterative ensemble Kalman method (IEnKM)~\cite{iglesias2013ensemble}, along with an ensemble-based regularizing scheme~\cite{iglesias2016regularizing}. Instead of directly sampling the entire posterior distribution, the Bayesian analysis formula in the IEnKM is derived under a Gaussian assumption. Specifically, by assuming that measurement noises $\bs{\sigma}_d$ obey an unbiased Gaussian distribution with a covariance $P_d$ and the underlying distribution of model predictions is also Gaussian with the mean $\mathbf{\bar{x}}$ and covariance $P_m$, the updated state (i.e., Bayesian analyzed state with a maximized posterior) $\mathbf{\hat{x}}$ can be expressed as,
\begin{equation}
\label{eq:update}
\mathbf{\hat{x}} = \mathbf{x} + P_mH^T(HP_mH^T + \alpha P_d)^{-1}(\mathbf{y} - H\mathbf{x}),
\end{equation}
where ${[\cdot]}^T$ denotes matrix transpose; $H$ is the matrix form of the observation projection function $\mathcal{H}(\cdot)$ mapping the full state $\mathbf{x}$ to the observed state $\mathbf{y}$; $\alpha$ is a control variable used for regularization described below. The Monte Carlo method is employed to estimate associated statistical information. Namely, the error covariance matrices $P_m$ and $P_d$ for the forward model predictions and observation data are estimated based on a number of samples. Therefore, potential non-Gaussian behavior and nonlinearity of the model can be considered by the ensemble-based estimations. Conceptually, to perform IEnKM-based data assimilation, there are three main steps: (1) prior sampling, (2) forward prediction (3) Bayesian analysis. These procedures are presented in the Fig.~\ref{fig:enkf}, and will be detailed below. The entire algorithm can be found in Appendix~\ref{sec:append}
\begin{figure}[!htbp]
	\centering
	\includegraphics[width=1.0\textwidth]{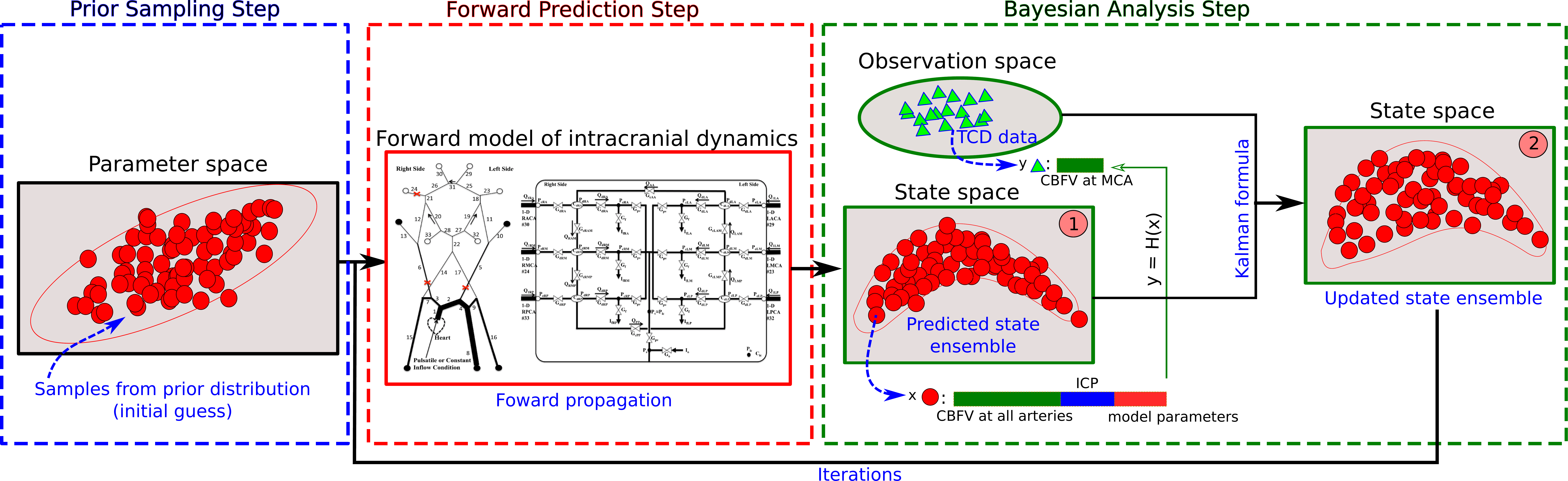}
	\caption{Schematic of the iterative ensemble Kalman method. (1) Initial ensemble is obtained by sampling the prior parameter space and (2) is propagated through the forward intracranial model. (3) The propagated state will be updated by assimilating TCD measurement data by Bayesian analysis. Steps (2) and (3) will be conducted iteratively until reaching the statistical convergence.}
	\label{fig:enkf}
\end{figure}

\textbf{Prior sampling:} The variations of model predictions (i.e., predicted CBFV and ICP) are induced by uncertainties in model parameters $\bs{\theta}$, initial physical state $\mathbf{z}_o$, and boundary conditions $\partial D$. To capture correlations among them, an augmented state vector $\mathbf{x} = [\mathbf{z}_o^T, \bs{\theta}^T, \partial D^T]^T$ is used in the data assimilation process. To begin, we sample the initial parameter space based on prior knowledge to represent the uncertainties in model parameters and initial/boundary conditions. The Latin hypercube sampling method~\cite{iman2008latin} is adopted to efficiently generate the initial state ensemble $\{\mathbf{x}_{j}\}^{N_s}_{j = 1}$, where $N_s$ represents number of samples. As shown in Fig.~\ref{fig:enkf}, each sample (red dot in the blue dashed box) represents one possible initial guess of the model inputs, i.e., model parameter set, initial and boundary conditions. 

\textbf{Forward prediction:} The uncertainties in model inputs lead to different predicted states by evaluating the forward intracranial model $N_s$ times. Namely, each initial sample in the parameter space corresponds to one possible physical state predicted by the forward model. The physical state vector consists of variables including ICP and CBFV and ABP at various vessels. Augmented with model parameters and boundary conditions, an ensemble of predicted states $\{\mathbf{x}_{j}\}^{N_s}_{j = 1}$ is obtained, which are represented by the red dots in the state space ``1'' in Fig.~\ref{fig:enkf}. Without incorporating any data, the forward prediction can be seen as an uncertainty propagation process, where variations of model inputs are propagated to the predicted state variables as propagated uncertainties. The predicted state ensemble represents a \emph{prior} estimation of the intracranial state.

\textbf{Bayesian analysis:} When data (i.e., TCD-based CBFV and/or ABP) are available, the predicted state ensemble can be updated based on the Eq.~\eqref{eq:update}. This is Bayesian analysis where both the physical variables and model parameters are updated by incorporating information from observation data. To calculate the analyzed state $\mathbf{\hat{x}}$, mean and error covariance information of predicted states and observation data is estimated by samples. The perturbed observation samples $\{\mathbf{y}_{j}\}^{N_s}_{j = 1}$ (blue triangles in Fig.~\ref{fig:enkf}) are obtained within the observation space spanned by measurement uncertainties and process errors. Note that only a very limited portion of the state is observed (e.g., CBFV at MCA), and thus the dimension of observation vector $\mathbf{y}$ is much smaller than that of the full state vector $\mathbf{x}$, as shown in Fig.~\ref{fig:enkf}. Finally, the updated state ensemble is in turn used as the initial ensemble in next iteration of the IEnKM. 

\textbf{Iterative regularization scheme:} The forward prediction and Bayesian analysis steps are conducted iteratively until a prescribed stopping criterion. To stabilize the Bayesian update and control the iterative process, a regularization scheme proposed in~\cite{iglesias2016regularizing} was adopted. Specifically, the control variable $\alpha$ in Eq.~\ref{eq:update} is calculated by the following sequence,
\begin{equation}
\alpha_{i+1} = 2^i\alpha_0,
\end{equation}
where $\alpha_0$ is an initial guess. Then, we chose $\alpha = \alpha_{N}$, where $N$ is the first integer such that,
\begin{equation}
\label{eq: alpha}
\alpha_N||P_d^{-1/2}(HP_mH^T + \alpha_NP_d)^{-1}(\mathbf{y}-H\mathbf{\bar{x}})||_2 \geq \rho||P_d^{-1/2}(\mathbf{y}-H\mathbf{\bar{x}})||_2,
\end{equation}
where $||\cdot||_2$ represents L2 norm, and $\rho$ is a constant parameter within an interval of $(0, 1)$. Larger $\rho$ indicates slowly decaying $\alpha$ and thus more regularization on the Bayesian analysis. The iteration is terminated whenever the normalized misfit between prediction and data is smaller than the noise level of the data, as shown by Eq.~\ref{eq:stop}. This regularization scheme can be derived as an approximation of the regularizing Levenberg-Marquardt scheme~\cite{more1978levenberg}, where the derivative of the forward operator and its adjoint are approximated using ensemble-based covariances. The details of associated derivations and proofs can be found in~\cite{iglesias2016regularizing}.

\section{Results}
\label{sec:result}
We first consider synthetic data to systematically explore the proposed framework. Our goal is to test if the assimilation of middle cerebral artery (MCA) blood flow velocity, which is readily accessible clinically, is sufficient to improve prediction of ICP in the model. Note, it is not obvious that assimilation of MCA CBFV data alone (and even if the data is noise-free) can lead to significant improvement in ICP prediction since the intracranial model is highly nonlinear, and MCA flow velocity has no direct relation to ICP. We then proceed to a more realistic application, using patient-specific TCD data measured clinically in patients suspected of having intracranial hypertension. These patients also had invasive ICP measurements performed that the model prediction can be compared against. 

Based on a parameter sensitivity analysis ({see Table.~\ref{tab:sensitivity}}) for the intracranial model by {the one-factor-at-a-time (OFAT) method}, we identified that the target perfusion flow rate parameters $q_n$ (see Eq.~\ref{eq:qn}) of the six arterial territories are important for both CBFV and ICP prediction. (Note, this does not necessarily imply CBFV is dominant in ICP prediction.) To improve identifiability of the problem, only the six primary parameters $q_n$ are inferred simultaneously along  with the hidden ICP state. Other parameters, which were deemed less important by the sensitivity analysis were determined offline from population-based calibrations conducted in previous studies~\cite{hu2007estimation,ursino2010model}. {The full set of primary parameters for both the forward intracranial model and the data assimilation process are given by Table~\ref{tab:parameter}.}

\begin{table}[htbp]
	\caption{{Sensitivity of primary model parameters to CBFV and ICP. Specifically, each parameter is uniformly perturbed by 20 percent of its baseline value, and the corresponding perturbations of CBFV and ICP are presented.}} 
	\label{tab:sensitivity}
	\small 
	\centering 
	\begin{tabular}{llllllllll} 
		\toprule[\heavyrulewidth]\toprule[\heavyrulewidth]
		Parameters &$r_{d0}$ & $\sigma_{e0}$ & $K_{\sigma}$ & $\sigma_{coll}$ & $\eta$ & $T_0$ & $t_{CA}$ & $G_{CA}$	& $q_n$\\
		\midrule
		CBFV & $1.25\%$ & $0.02\%$ & $0.05\%$ & $0.60\%$ & $0.28\%$ & $3.58\%$ & $0.01\%$ & $0.16\%$ & $35.57\%$\\
		ICP &  $0.90\%$ & $0.03\%$ & $0.04\%$ & $0.52\%$ & $0.03\%$ & $2.89\%$ & $0.02\%$ & $0.05\%$ & $30.83\%$\\
		\bottomrule[\heavyrulewidth] 
	\end{tabular}
\end{table}

\begin{table}[htbp]
	\caption{{Primary parameters of forward model and data assimilation}} 
	\label{tab:parameter}
	\small 
	\centering 
	\begin{tabular}{lll} 
		\toprule[\heavyrulewidth]\toprule[\heavyrulewidth]
		\multicolumn{3}{c}{Baseline values of forward intracranial model}\\
		\midrule
		$r_{d0} = 0.015\ \mathrm{cm}$ & $\sigma_{e0} = 0.1425\ \mathrm{cm}$ & $K_{\sigma} = 10.0$\\
		$\sigma_{coll} = 62.79\ \mathrm{mm\ Hg}$ &  $\eta = 232\ \mathrm{mm\ Hg}$ & $T_0 = 2.16\ \mathrm{mm\ Hg\ cm}$\\
		$r_t = 0.018\ \mathrm{cm}$ &  $r_m = 0.027\ \mathrm{cm}$ & $t_{CA} = 10\ \mathrm{s}$  \\
		$G_{CA} = 10\ \mathrm{mm\ Hg}^{-1}$ &  \multicolumn{2}{c}{$q_{n} = 2.2$ (MCAs), $1.48$ (ACAs), $1.14$ (PCAs) $\mathrm{ml\ s}^{-1}$}\\
		\toprule[\heavyrulewidth]
		\multicolumn{3}{c}{Parameters of data assimilation (IEnKM)}\\
		\midrule
		prior uncertaintiy & \multicolumn{2}{c}{$20 \%$ uniformly random perturbation}\\
		number of samples $N_s$ &  \multicolumn{2}{c}{$20$}\\
		regularization parameters &  \multicolumn{2}{c}{$\rho = 0.6, \alpha_0 = 1$}\\
		\bottomrule[\heavyrulewidth] 
	\end{tabular}
\end{table}

\subsection{Verification with synthetic data}
The intracranial model was run using an arbitrary but physiologic set of model parameters as the ``ground truth''. That is, instead of data coming from a patient, data comes from the model run with a hidden parameter set. The six unknown ``true'' target flow rate parameters are shown as black lines in Fig.~\ref{fig:syn_noErr_para}. Synthetic TCD data was obtained by ``measuring'' CBFV at the left and right MCAs, with and without artificial random noises added. Then all simulated information is discarded, except the measured MCA CBFV data, and the ICP which was blinded and reserved as the ``ground truth'' to later compare against. {To determine the sample size sufficient for an accurate mean estimation, data assimilation using $N_s = 20, 50$, and $100$ samples was conducted and the expectations of posterior ICP estimations are compared. The results showed that the difference among these cases was less $2\%$. Therefore, twenty samples ($N_s = 20$) were adopted in the following numerical cases.} {Note that the term ``sample'' represents one of the randomly perturbed forward simulations in IEnKM, while the term ``data'' refers to the noninvasive measurements within this paper.}

\subsubsection{Noise-free CBFV data}

\begin{figure}[htbp]
	\centering 
	\includegraphics[width=1.0\textwidth]{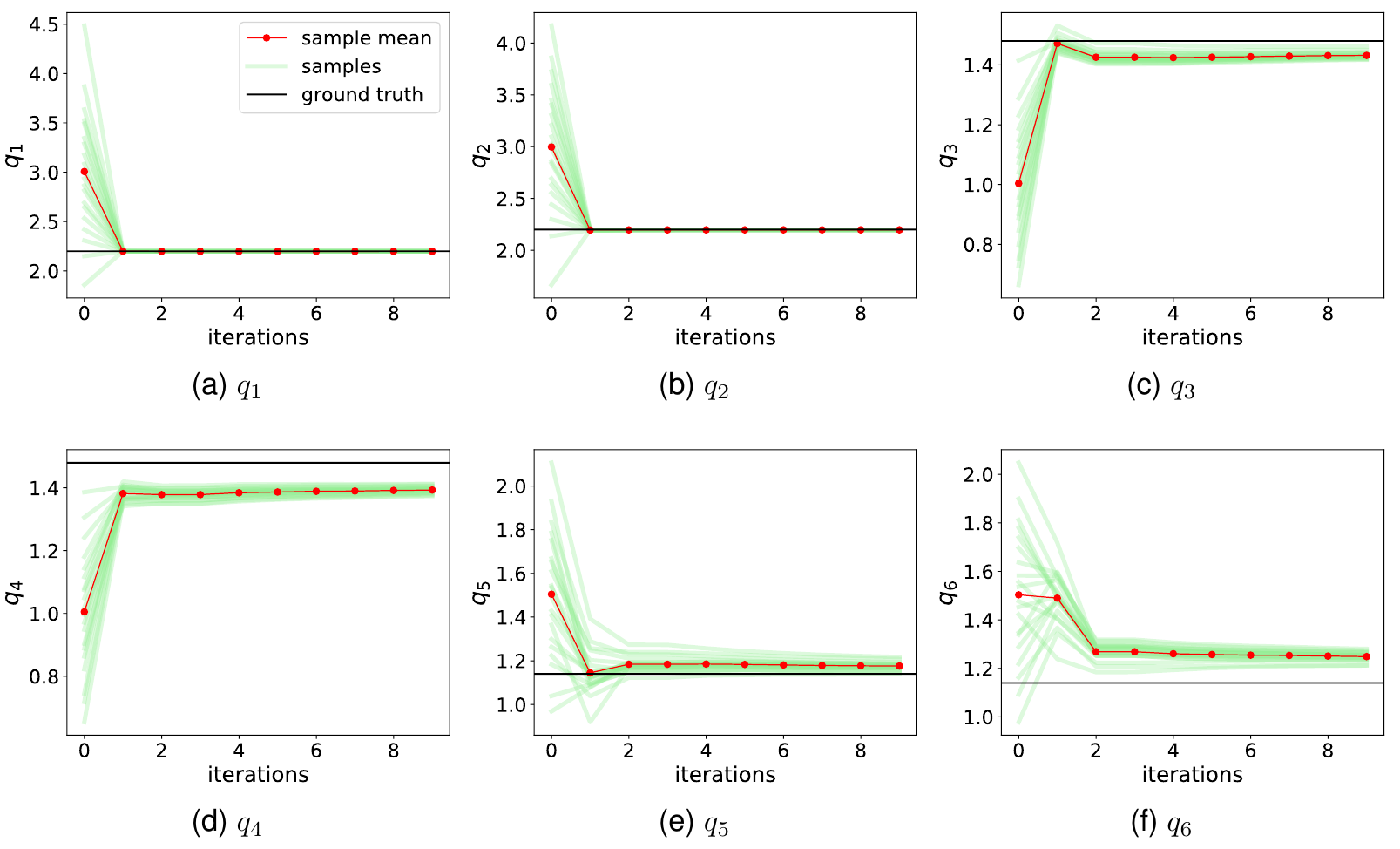}
	\caption{Iteration histories of unknown parameters (i.e., target flow rates $q_n$, $n = 1 \cdots, 6$) by assimilating noise-free synthetic TCD data. The prior ensemble of each parameter is biased from the respect truth.}
	\label{fig:syn_noErr_para}
\end{figure}
We first consider the case in which the MCA CBFV measurements are noise-free. Figure~\ref{fig:syn_noErr_para} shows DA iteration histories of the target flow rate parameters $q_n, n = 1, \cdots, 6$, which are added to the extended state vector and inferred during the DA process. All the samples (light green lines) are scattered initially (iteration step 0), representing the prior distributions of the parameters. Each ensemble mean at iteration step 0 (first red dot) can be seen as a best prior guess for the respective parameter; it is observed that these prior guesses deviate significantly from the respective ground truths (black lines), representing a typical biased prior estimation. The bias is highly relevant in real-world applications, since it is almost impossible to guarantee an unbiased prior (initial guess for a \emph{de novo} patient). However, after assimilating the synthetic TCD velocity ``measurements'' at the MCAs, the unknown parameters are well recovered within a few iterations and uncertainty is largely reduced. As expected, CBFV data in the MCAs is most informative to MCA-related target flow rates ($q_1$ and $q_2$), which converge exactly to the truth, whereas slight discrepancies remain for ACA and PCA territory perfusion flow rates ($q_3$ -- $q_6$).  

\begin{figure}[htbp]
	\centering
	\includegraphics[width=1.0\textwidth]{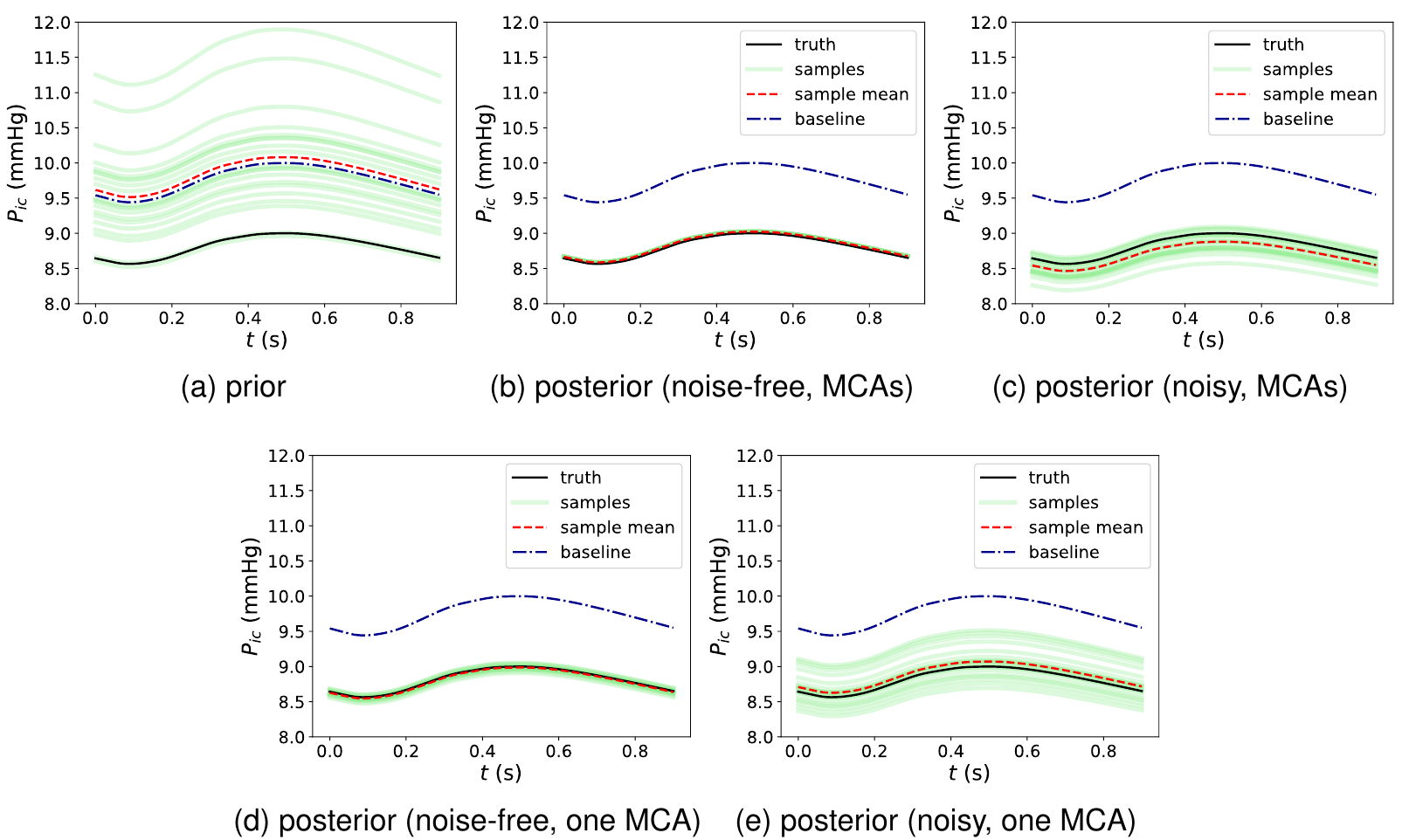} 
	\caption{{Comparison of (a) prior ICP prediction and posterior ICP predictions following (b) noise-free, two MCAs, (c) noisy, two MCAs, (d) noise-free, right MCA and (e) noisy, right MCA CBFV data assimilations.}}
	\label{fig:syn_noErr_icp}
\end{figure}
We next consider the hidden state of most interest, ICP, which is less directly related to CBFV than the hidden parameters $q_n$. The comparison of prior and posterior estimations for ICP are presented in Fig.~\ref{fig:syn_noErr_icp}. In Fig.~\ref{fig:syn_noErr_icp}a, it can be observed that the prior samples of ICP prediction are scattered from around $8.5$ mmHg to $12.0$ mmHg due to physiologic perturbations of the initial state and parameters due to prior epistemic uncertainty. Similarly, the mean (red dashed) of the prior ICP ensemble is biased from the truth (black solid). Fig.~\ref{fig:syn_noErr_icp}b displays the convergence of each sample following assimilation of the MCA CBFV data, demonstrating that all the posterior samples {converge} to the truth after the regularizing iterative ensemble Kalman DA method is applied with very low uncertainty and expected value very close to the true value. We note that the other hidden physical states, including flow velocity and pressure at unobserved arteries, were also similarly recovered via this framework; since the prediction performance is similar to that shown here for ICP, those results are omitted.

{Assimilating data from both MCAs was considered in the synthetic tests since such simultaneous measurements can, in theory, be obtained clinically. Alternatively, we also consider data from only one MCA, which is more consistent with the clinical cases considered below in Section~\ref{sec:result_clinical}. Results from the synthetic tests with only the right MCA CBFV data assimilated are presented in Fig.~\ref{fig:syn_noErr_icp}, with the results from noise-free data in panel (d) and from noisy data in panel (e). As shown, the posterior mean of ICP maintains close consistency with the true ICP. However, the posterior sample scattering is slightly higher compared to the case with data from both MCAs assimilated (Fig.~\ref{fig:syn_noErr_icp}, panels b and c). This demonstrates that the epistemic uncertainties resulting from the lack of data can be reasonably considered in the current Bayesian framework.}

\subsubsection{Noisy CBFV data}

We next consider corrupting the synthetic MCA CBFV data with $10\%$ Gaussian random noise to represent measurement error, and in addition a $10\%$ process error is considered to account for potential model-form uncertainties. The combination of the measurement error and process error are reflected by the data error covariance matrix $P_d$~\cite{dennis2006estimating}. We focus here on our ultimate target of ICP. Figure~\ref{fig:syn_noErr_icp}c displays the ICP posterior estimation following assimilation of the noisy MCA CBFV data. It is clear that all ICP samples, which as above demonstrate high scatter in the prior estimation, converge toward the true signal by incorporating the (now noisy) CBFV data, and that the associated posterior uncertainties are largely reduced. However, compared to the results of the noise-free case as shown in Fig.~\ref{fig:syn_noErr_icp}b, where all posterior ICP samples {converge} to the truth, the posterior ICP samples in Fig.~\ref{fig:syn_noErr_icp}c display some scatter, or posterior uncertainty. Nonetheless, all samples and the expectation are close to the truth. {Moreover, if only the data from one MCA is assimilated, a higher posterior uncertainty (i.e., higher sample scatter) is observed in Fig.~\ref{fig:syn_noErr_icp}e, which is expected due to the reduction of data.}

\begin{figure}[htbp]
	\centering
	\includegraphics[width=1.0\textwidth]{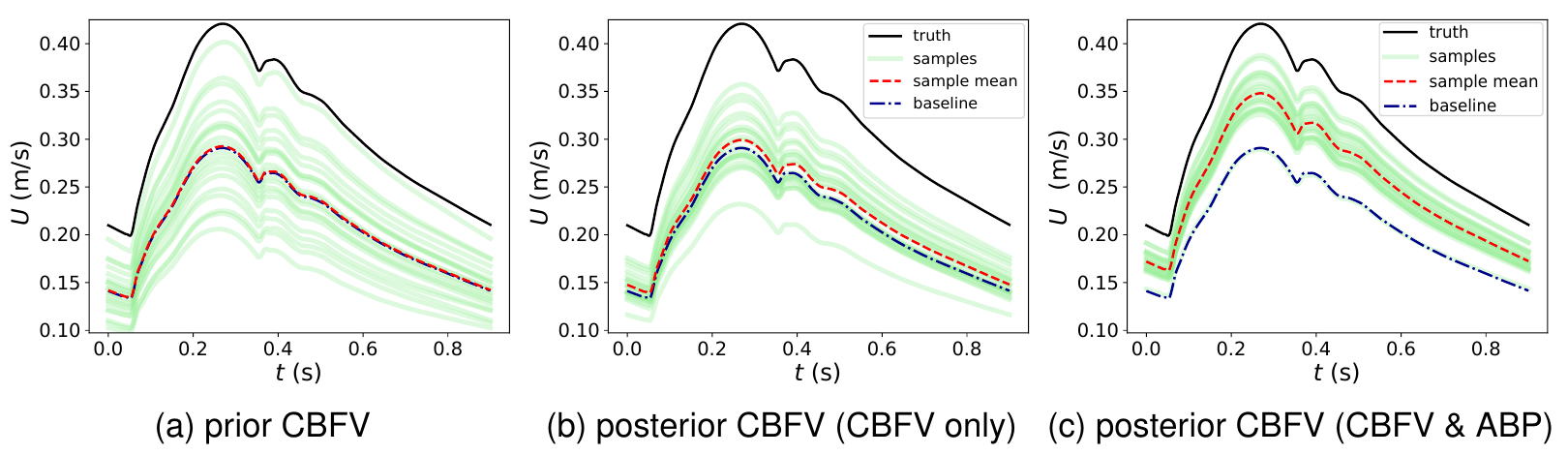} 
	\caption{Comparison of prior and posterior predictions of CBFV at right ACA following noisy synthetic data assimilation. In (b) CBFV data at two MCAs with $10\%$ Gaussian noises are assimilated; In (c) Both CBFV and ABP data at two MCAs with $10\%$ Gaussian noises are assimilated.}
	\label{fig:syn_10pErr_u}
\end{figure}

Interestingly, in contrast to the ideal noise-free case above where all the hidden states are essentially recovered exactly, the posterior estimations of CBFV at unobserved vessels are not significantly improved by assimilating the noisy data. For example, Figures~\ref{fig:syn_10pErr_u}a and~\ref{fig:syn_10pErr_u}b  show the prior and posterior ensembles of CBFV at the right ACA, where the improvement of the posterior sample mean is not notable compared to the prior, and large uncertainties remain in the posterior estimation. This indicates that CBFV data at the MCA is not necessarily informative to CBFV at other arteries, and exemplifies that it is not trivial to predict which (presumably measurable) states will be significantly informative of other (presumably hidden) states. 

\subsubsection{Assimilating additional ABP data}

The cases showed above only utilize CBFV data in the MCAs, which is commonly measurable by TCD ultrasonography. Other than the TCD-based CBFV data, systemic ABP measurements are also among the data available bedside in routine clinical practice. Therefore, it is also interesting to investigate whether the posterior predictions can be further improved by assimilating both ABP and CBFV measurements simultaneously. We conducted another experiment with the same set up as above, except using both CBFV and ABP data sampled from two MCAs. By additionally incorporating the ABP data, performance of the ICP prediction remained excellent, and additionally posterior estimations of CBFV at unobserved arteries were notably improved. For example, Figure~\ref{fig:syn_10pErr_u}c shows the posterior ensemble of CBFV at the right ACA by assimilating both CBFV and ABP data at the MCAs. All the samples are corrected toward the ground truth and the posterior mean is significantly improved compared to the results displayed in Fig.~\ref{fig:syn_10pErr_u}b. Moreover, sample scattering is also relatively smaller. Note that the ABP assimilated in our model was arterial pressure at the MCA whereas systemic ABP data are typically measured at the radial artery, which differ in time and waveform. A correction algorithm proposed by Kashif et al.~\cite{kashif2012model} needs to be employed to obtain an approximation of ABP at MCAs when systemic ABP data measured from the radial artery are used.

\subsection{Validation against clinical data}
\label{sec:result_clinical}
Preliminary clinical application of the proposed framework was also investigated. TCD measurements were obtained in patients that had invasive ICP measured. Note, only right MCA CBFV was assimilated in these validation studies, compared to both left and right MCA data being assimilated in the synthetic cases above. Generally, the prior model parameters and overall framework were the same as in the synthetic test cases, except the observation data to be assimilated. However, the main difference here from synthetic cases is not just that real versus synthetic TCD data was used, but that the ICP we compare against was measured from actual patients, and not the computational model. 

The TCD and ICP data were acquired by the protocol approved by the UCLA Internal Review Board and the full dataset was reported in~\cite{kim2013noninvasive}. In this study, we focus on patients with a homeostatic intracranial system, where the ICP/CBFV waveform is assumed to have reproducible features at similar mean levels~\cite{hu2012steady}. We investigate two patients with approximate homeostatic TCD and ICP signals. The two patients are referred to as P1 and P2. Patient P1 was a 55 years-old female and treated for aneurysmal subarachnoid hemorrhage (aSAH), while patient P2 was a 47 years-old male and treated for traumatic brain injury (TBI). The ICP signals for both patients were obtained through external ventricle drainage (EVD). 
\begin{figure}[htbp]
	\centering 
	\includegraphics[width=0.8\textwidth]{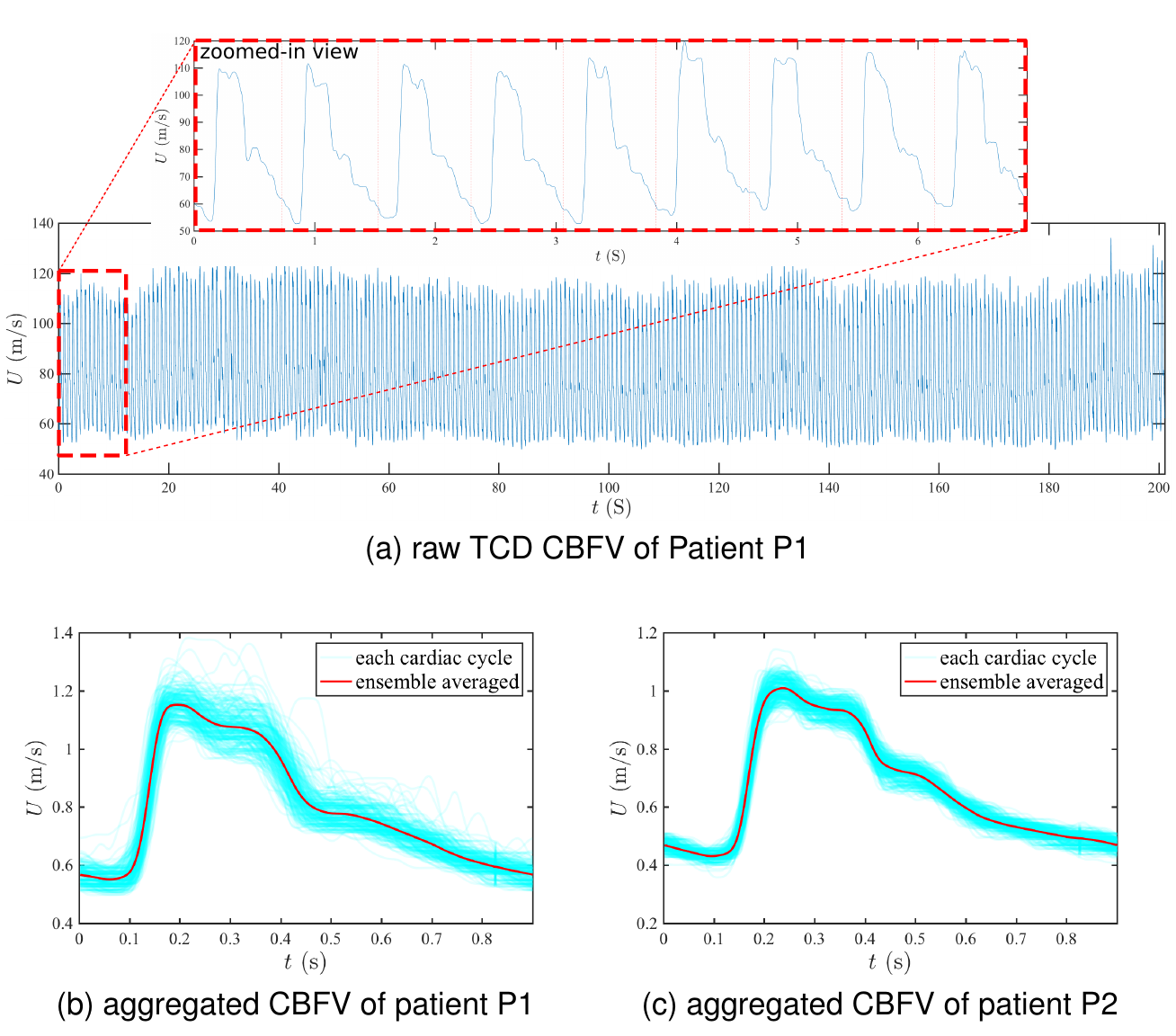}
	\caption{TCD CBFV data. (a) Raw TCD-based CBFV signals at right MCA of the patient P1 over 260 cardiac cycles. (b-c) Aggregated CBFV pulses at the right MCA and its ensemble averaged for (b) patient P1 and (c) patient P2. }
	\label{fig:real_raw_TCD}
\end{figure}
Figure~\ref{fig:real_raw_TCD}a displays the raw TCD-measured CBFV signal at right MCA for patient P1 over 260 cardiac cycles. The mean CBFV level approximately remains the same, and waveform features are also similar cycle to cycle, as shown in the zoomed-in view of Fig.~\ref{fig:real_raw_TCD}a. Similar steady features are also observed in the corresponding ICP signal and for the MCA CBFV and ICP data for patient P2. 

Based on the quasi-steady nature of the signals, pulses over the 260 cardiac cycles of raw CBFV signal data were aggregated and the ensemble average was computed. Figures~\ref{fig:real_raw_TCD}b and~\ref{fig:real_raw_TCD}c show the CBFV data of patients P1 and P2 where all pulses from the raw signal are plotted within one cardiac cycle, and the ensemble-averaged pulse is plotted by a bold red line. The shape of the mean pulse is triphasic (i.e., having three peaks) for both patients, which is a commonly observed feature for both CBFV and ICP waveforms~\cite{hu2009morphological, piper1990systems, piper1993experimental}. Although the waveforms of the MCA CBFV between the two patients are similar, the mean CBFV level of patient P1 is slightly larger than that of patient P2. 

\begin{figure}[htbp]
	\centering
	\includegraphics[width=0.66\textwidth]{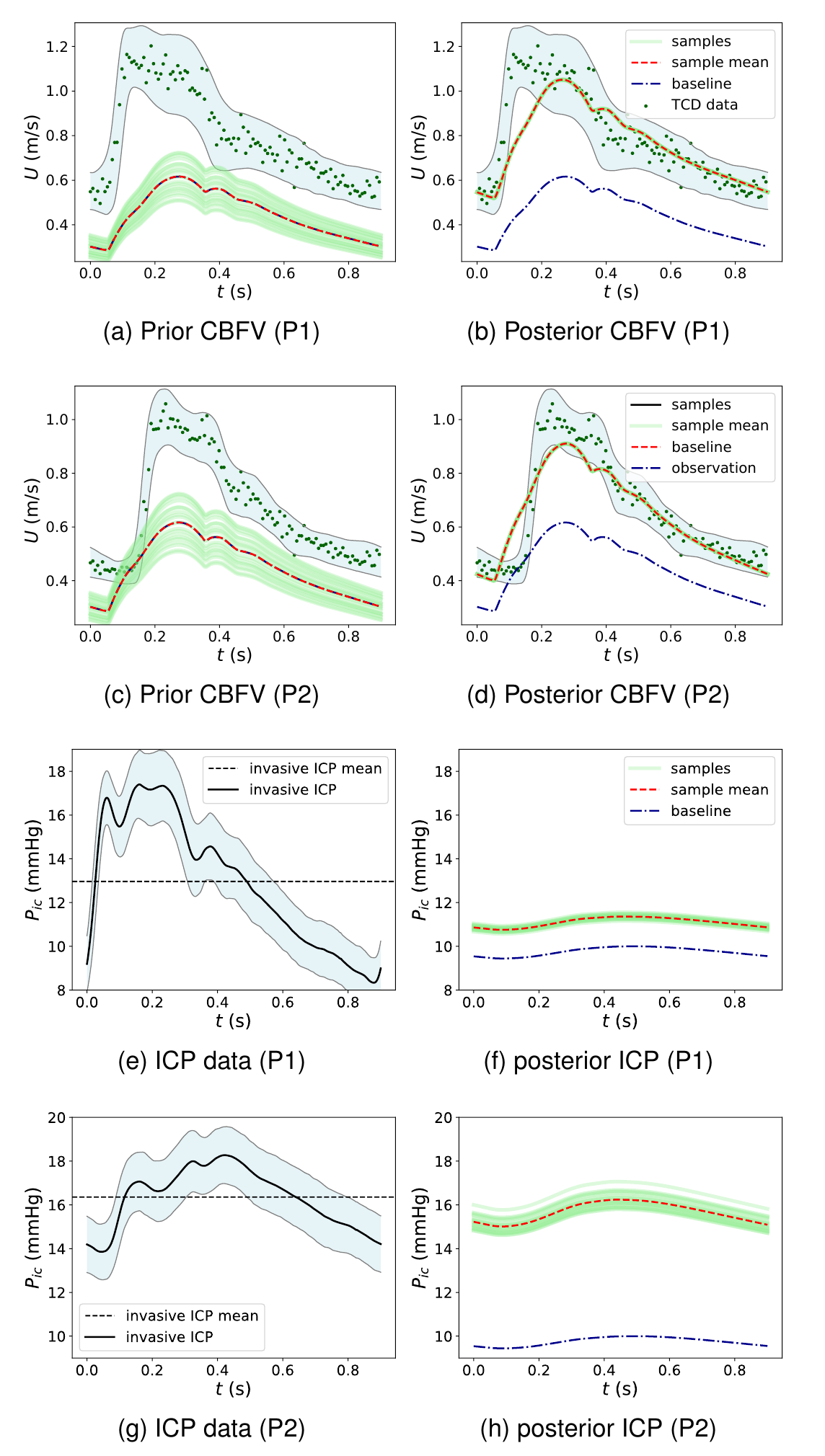} 
	\caption{CBFV predictions (a-d) and ICP predictions (e-h) following assimilation of TCD CBFV measurements.}
	\label{fig:real_u_p1}
\end{figure}

The scattering of the pulse data represents uncertainties introduced by TCD measurement errors, respiratory effects, and physiological deviations from the steady-state assumption. These uncertainties are treated as data uncertainties in the assimilation process and are estimated based on the pulse history. That is, instead of assimilating the ensemble average, which would effectively ignore this uncertainty, the statistical distribution of the measured CBFV data was used to sample data for assimilation. For example, Fig.~\ref{fig:real_u_p1}a for patient P1 (and in Fig.~\ref{fig:real_u_p1}c for patient P2) displays the uncertainty interval of TCD data (light blue region) and sampled data for assimilation (green dots). As for the synthetic cases above, the CBFV data is assimilated to calibrate target flow rate parameters, which were deemed to be informative of ICP via the OFAT sensitivity analysis. 
We first compare the ability of the model to match the measured TCD by simulating the model with the inferred target flow rates for each patient. Without data assimilation, the prior predictions of RMCA CBFV are highly scattered and biased, and the mean CBFVs are largely underestimated for both patients, as shown in Figs.~\ref{fig:real_u_p1}a and~\ref{fig:real_u_p1}c. Following data assimilation, the model parameters (i.e., target flow rates to each territory) appear well inferred as the corresponding RMCA CBFV posterior prediction is significantly improved when compared with the TCD data and has less uncertainty, as shown in Figs.~\ref{fig:real_u_p1}b and~\ref{fig:real_u_p1}d. Note that all the posterior samples are mostly {converged} to the sample mean curve (red dashed line) quickly. The reason for this is that the data uncertainty level is relatively large compared to the perturbation of simulations and thus the stopping criteria can be satisfied within only a few DA iterations. Although the posterior CBFV pulse agrees well with TCD data and mostly falls inside the data uncertainty region, some discrepancy can be observed at the beginning of the CBFV pulse (``early systole") in both patients. This is likely due to the model-form error as discussed below.

Finally, we investigate the prediction of ICP for the two patients in comparison to the invasive ICP data measured from the lateral ventricles in each patient. Similar to the TCD data, the raw ICP pulses are aggregated to compute an ensemble average and statistical distribution for uncertainty. Figures~\ref{fig:real_u_p1}e and~\ref{fig:real_u_p1}g show the ensemble-averaged pulse for the invasive ICP measurements for patients P1 and P2, while Figs.~\ref{fig:real_u_p1}f and~\ref{fig:real_u_p1}h display the noninvasive ICP predictions obtained from our data assimilation framework. For both patients, the baseline (prior mean) ICP prediction substantially under-predicts the measured mean ICP. However, after assimilating the TCD measurements, the posterior ICP predictions of both patients (red curve) increase significantly, with a mean value reasonably close to the mean invasive ICP measurements for each respective patient. Specifically, for patient P1 (Figs.~\ref{fig:real_u_p1}e and~\ref{fig:real_u_p1}f), the mean ICP measurement is 12.8 mmHg and the mean of TCD-augmented ICP prediction is 11.2 mmHg. And for patient P2 (Figs.~\ref{fig:real_u_p1}g and~\ref{fig:real_u_p1}h), the mean ICP measurement is 16.2 mmHg and the mean of TCD-augmented ICP prediction is 15.9 mmHg. For both patients, the prediction error in mean ICP is within 2 mmHg, which is the clinically-accepted ICP error standard~\cite{zhang2017invasive}. Moreover, it is interesting to note that the posterior ICP samples exhibit relative scatter although the corresponding CBFV samples are mostly {converged}. This is because no direct ICP measurements are used in the assimilation and thus relatively larger epistemic uncertainties are expected. Although mean ICP appears well predicted and significantly improved using the framework proposed herein, the predicted ICP waveform shape significantly differs from the measured triphasic shape. This discrepancy is discussed below.

\section{Discussion}
\label{sec:dis}
We have presented a data-augmented, theory-based modeling approach for noninvasive intracranial pressure estimation, based on a multiscale intracranial model and assimilation of clinically-available TCD CBFV data. A regularizing iterative ensemble Kalman method is employed for fusing the computational model with measurement data. The proposed framework has been examined through both synthetic tests and tests with actual patient data, both of which demonstrated that the presented assimilation procedure was able to significantly improve mean ICP prediction.  

The tests using synthetic data were conducted to verify implementations of the framework and analyze the identifiability of the unknown parameters and hidden variables. When both the forward intracranial model and measurement data are precise (i.e., no error in the model or synthetic data), all the unknown parameters and hidden states including unobserved CBFV and ICP can be precisely recovered by only assimilating CBFV data at the MCAs and associated uncertainties due to prior perturbations in model parameters can be nearly eliminated. For a more realistic condition, where both the forward model and measurement data were made imprecise through the introduction of a 10\% error, strong performance of ICP prediction was maintained. Namely, the posterior mean of the ICP agreed well with the ground truth, however uncertainty in the posterior ICP prediction increased due to the uncertainties introduced by the model inadequacy and measurement noise. Nonetheless, the results demonstrate that overall the ICP prediction can be significantly improved by incorporating noninvasive CBFV data, and uncertainties associated with the data and model can be naturally considered within the Bayesian framework presented.

Based on the results of the synthetic tests, CBFV data at MCAs were demonstrated to be informative to ICP predictions. However, they appeared less informative to CBFV at other unobserved arterials, e.g., ACAs, when data were corrupted by random noise. Although ICP can be accurately predicted, the improvement of posterior predictions of CBFV at unobserved arterials was not remarkable and uncertainties remained considerable after data assimilation. However, by additionally incorporating ABP data along with CBFV data at MCAs, predictions of CBFV at ACAs were significantly improved. These results indicate that assimilating more independent information can further enhance hidden state estimation and reduce associated epistemic uncertainties. The assimilations of multiple noninvasive signals is something to be further explored to broaden or further improve the utility of this framework. 

Actual TCD and invasive ICP measurements from two patients with approximate homeostatic intracranial dynamics were used to examine the feasibility of the proposed approach toward clinical application. The performance of the proposed approach in these patient-based cases was promising. The data assimilation procedure led to significant improvement in mean ICP prediction, with a posterior estimate of mean ICP close to the invasively measured mean value and within the current clinical standard for ICP error. These results indicate that noninvasive ICP prediction can be informed by CBFV data, and implies that the perfusion blood flow distribution among the six major vascular territories appears to be a significant factor in steady-state ICP dynamics. {While the focus of this paper is the theoretic basis and methodology of the data-augmented nICP framework, these clinical comparisons, although limited, demonstrate feasibility of the proposed approach. Nonetheless, additional validations are needed to properly establish clinical viability of this approach.} 

It should be noted that although mean ICP prediction matched reasonably with that from invasive measurement, the shape of the measured ICP waveform was not well replicated. This may be expected for several reasons. First, the shape of the posterior prediction for the MCA CBFV waveform differed (albeit to less degree) from the measured waveform (Figs.~\ref{fig:real_u_p1}b and~\ref{fig:real_u_p1}d). {This is potentially due to the simplified sinusoidal waveform used at the aortic root, which essentially drives waveform dynamics to the rest of the model. While it is possible to impose a more physiologic waveform at the aortic root, ideally these dynamics should arise naturally from the model assuming that an appropriate ``unadulterated" waveform can be imposed, since the measured aortic waveform already contains reflected waves, which would be confounded by reflections generated from the 1D network model.} Second, ICP modeling herein was highly simplified, which likely contributes to the damped dynamics of the ICP waveform. In our model, intracranial pressure and CSF was modeled as spatially uniform and shared by the six distal vascular beds. CFS (and hence ICP) dynamics were governed by simple conductances at the capillary outlets and venous return. This is a significant simplification and in reality ICP dynamics is likely influenced by the multi-ventricular flow of CSF and dynamic coupling with brain tissue and different arterial territories. Hence, it is expected that the forward model should be geared toward improved ICP dynamics modeling by considering expanded modeling of the CFS circulation and dynamic coupling with the brain and other tissues. (Indeed, the computed and measured CBFV waveforms agreed much more closely, even without calibration, as the intracranial model employed was more hemodynamics-oriented.) This is a significant undertaking and will be pursued in a separate work, however, it is important to note that mean ICP is generally used clinically for diagnosis of intracranial hypertension. Nonetheless, it is expected that improved modeling of CSF and ICP dynamics will improve the predictive capabilities of even mean ICP, and moreover emerging research~\cite{connolly2015characterization,arroyo2016characterization} is recognizing the importance of ICP waveform analysis for diagnosis and differentiation of cerebral pathologies and treatment management. 

{In regards to numerical implementation, the majority of the forward intracranial model dynamics is implemented in C++, where the 1D distributed network is solved using an in-house finite volume solver and the LP intracranial portion is solved using an open-source ODE solver of a C++ library SUNDIALS~\cite{serban2005cvodes}. The in-house data assimilation solver, i.e., IEnKM, was implemented in Python.} {The computational cost of this implementation mainly depends on the number of samples used for Kalman updates, since each sample involves a forward simulation. As mentioned above, $N_s = 20$ samples are used in this work and approximately three to five iterations are needed to achieve statistical convergence. Therefore, each data assimilation case may involve approximately $100$ forward model evaluations, which entails running the model until it reaches homeostatic intracranial state. If starting from the steady state of the baseline case, each perturbed case typically converges in less than five cardiac cycles. On a single CPU core, it takes about 40 seconds to simulate one cardiac cycle. However, the propagation of case ensemble can be done in parallel. In this work, a dual-processor with 20 CPU cores was used and thus each data assimilation case took approximately 15 minutes. However, computational efficiency has not yet been a focus, since the objective of this work has been to explore feasibility.}

\appendix
\label{sec:append}
\section{Algorithm: regularizing iterative ensemble Kalman method}
\noindent\textbf{Prior sampling}: Use Latin hypercube sampling method to generate the prior state ensemble $\{\mathbf{x}_j^{(0)}\}_{j=1}^{N_s}$, where $\mathbf{x}_j$ is $j^{\mathrm{th}}$ sample of the augmented state, including major arterials' CBFV and ABP, ICP, and unknown parameters. Let $\rho \in (0, 1)$ and $\tau = 1/\rho$.

\noindent\textbf{For} $n = 1 : n_{max}$
\begin{enumerate}
	\item \textbf{Forward prediction:}\\
	(a) Evaluate the forward intracranial model with the initial physical state, boundary conditions, and model parameters, which are updated in the last iteration. Namely, the analyzed state $\mathbf{\hat{x}}_j^{(n)}$ at iteration step $n$ is propagated through the forward model $\mathcal{F}$ at $(n+1)^{\mathrm{th}}$ iteration,
	\begin{equation}
	\mathbf{x}_j^{(n+1)} = \mathcal{F}(\mathbf{\hat{x}}_j^{n}).
	\end{equation}
	
	(b) Obtain the perturbed ensemble of observation data $\{\mathbf{y}_j^{(0)}\}_{j=1}^{N_s}$ based on the data uncertainty level $\sigma_d$.
	
	(c) Calculate statistical information of predicted state and observation data. We first calculate the sample means of state and data as,
	\begin{eqnarray}
	\mathbf{\bar{x}}^{(n+1)} = \frac{1}{N_s}\sum_{j=1}^{N_s}\mathbf{x}^{(n+1)}_j\\
	\mathbf{\bar{y}}^{(n+1)} = \frac{1}{N_s}\sum_{j=1}^{N_s}\mathbf{y}^{(n+1)}_j.
	\end{eqnarray}
	The error covariances of the predicted state and observation data can then be obtained,
	\begin{eqnarray}
	P_m^{(n+1)} = \frac{1}{N_s - 1}\sum_{j=1}^{N}(\mathbf{x}^{(n+1)}_j - \mathbf{\bar{x}})(\mathbf{x}^{(n+1)}_j - \mathbf{\bar{x}}^{(n+1)})^T\\
	P_d^{(n+1)} = \frac{1}{N_s - 1}\sum_{j=1}^{N}(\mathbf{y}^{(n+1)}_j - \mathbf{\bar{y}})(\mathbf{y}^{(n+1)}_j - \mathbf{\bar{y}}^{(n+1)})^T
	\end{eqnarray}
	\item \textbf{Regularizing Bayesian analysis:}\\
	(a) Calculate the control variable $\alpha_i^{(n+1)}$ by following sequence,
	\begin{equation}
	\alpha_i^{(n+1)} = 2^{i+1}\alpha_0,
	\end{equation}
	where an initial guess of $\alpha_0 = 1$ is used in this work. The $\alpha^{n+1}$ is obtained as $\alpha^{(n+1)} \equiv \alpha^{(n+1)}_{N}$, where $N$ is the first integer when the inequality defined by Eq.~\ref{eq: alpha} is satisfied. 
	(b) Compute regularized Kalman gain matrix as,
	\begin{equation}
	K^{(n+1)} = P_m^{(n+1)}H^T(HP_m^{(n+1)}H^T + \alpha^{(n+1)}P_d^{(n+1)})^{-1},
	\end{equation}
	(c) Update each state sample as follows,
	\begin{equation}
	\mathbf{\hat{x}}^{(n+1)} _j= \mathbf{x}^{(n+1)} _j + K^{(n+1)}(\mathbf{\bar{y}}^{(n+1)} - H\mathbf{x}^{(n+1)} _j),
	\end{equation}
	
	\item \textbf{Stopping criteria:}\\
	If
	\begin{equation}
	\label{eq:stop}
	||{P_d^{(n+1)}}^{-1/2}(\mathbf{y}^{(n+1)} - H\mathbf{\bar{x}}^{(n+1)})|| \leq \tau\sigma_d,
	\end{equation}
	then, stop the iteration. $\sigma_d$ represents noise level of observation data. 
	
\end{enumerate}

\section*{Acknowledgments}
JXW would like to thank J. Pyne and J. Wu for helpful discussions. The authors also thank the anonymous reviewers for their comments and suggestions, which helped improve the quality and clarity of the manuscript.

\bibliographystyle{abme}


\end{document}